\documentclass[aps,prl,showpacs,notitlepage,nofootinbib,superscriptaddress,floatfix,showkeys,onecolumn,longbibliography]{revtex4-1}

\usepackage[table]{xcolor} 
\usepackage{geometry}

\usepackage{siunitx}
\usepackage{blindtext}
\usepackage{hyperref}
\usepackage{amsmath,amssymb}
\usepackage{float}
\usepackage{microtype}
\usepackage{slashed}
\usepackage{graphicx}
\usepackage{bm}
\usepackage{latexsym}
\usepackage{epsfig}
\usepackage{psfrag}
\usepackage{color}
\usepackage{subfigure}
\usepackage[section]{placeins}
\usepackage{array}   
\usepackage{colortbl} 
\usepackage{easyReview}
\usepackage{tcolorbox}
\usepackage{xfrac}
\usepackage{booktabs}

\usepackage{array} 

\def\e1i{\epsilon_{1\mathrm{i}}}

\allowdisplaybreaks[1]

\definecolor{customgreen}{HTML}{c7efcf}
\definecolor{customblue}{HTML}{9cafb7}

\smallskipamount

\begin{document}

\title{Implications of Photon Mass:  Vortextrap Magnetization of Black Holes}

\author{Gia Dvali}
\affiliation{Arnold Sommerfeld Center,  Ludwig Maximilians University, Theresienstra{\ss}e 37, 80333, Germany}
\affiliation{Max Planck Institute for Physics, 
Boltzmannstra{\ss}e 8, 85748 Garching Munich, Germany}

\author{Zaza N. Osmanov}\affiliation{School of Physics, Free University of Tbilisi, 0183, Tbilisi, Georgia}

\author{Michael Zantedeschi}		
\affiliation{INFN Sezione di Pisa, Polo Fibonacci, Largo B. Pontecorvo 3, 56127 Pisa, Italy}

\begin{abstract}
We discuss certain astrophysical implications of the photon mass. 
 It offers a new mechanism of black hole magnetization, described as   ``vortextrap magnetization"  (VTM), which can generate  a near-saturated magnetic field in 
astrophysical black holes. The extreme magnetic field is provided by a large number of Nielsen-Olesen type 
vortex lines piercing a black hole.  In massive photon scenario the galactic magnetic field 
is a densely populated forest of overlapping  magnetic flux tubes. 
 These get trapped and collected by a black hole over a  cosmological time-scale.
    The VTM mechanism neatly fits  supermassive black holes with  
 sizes  matching the  phenomenologically-acceptable values of the photon mass, 
 and  has implications for magnetic-field based particle acceleration.  
  Even in absence of surrounding plasma, the near-saturated magnetic field is expected to result into an intense electromagnetic radiation as well as gravitational waves  in black hole mergers.   
  We provide a numerical simulation of the VTM phenomenon in a prototype system.   
\end{abstract}

\maketitle

\begingroup
\renewcommand{\thefootnote}{\fnsymbol{footnote}}
  \footnotetext[1]{\href{mailto:gdvali@mpp.mpg.de}{gdvali@mpp.mpg.de}}
  \footnotetext[3]{\href{mailto:z.osmanov@freeuni.edu.ge}{z.osmanov@freeuni.edu.ge}}
  \footnotetext[4]{\href{mailto:michael.zantedeschi@pi.infn.it}{michael.zantedeschi@pi.infn.it}}
\endgroup

\section{Introduction}

 The purpose of the present paper is to point out  
certain astrophysical implications of a non-zero 
 mass of the photon. 
  The mass of photon, $m_{\gamma}$, is a basic microscopic parameter of nature with an uncertain value.  Needless to say, any evidence 
 of non-zero $m_{\gamma}$ would be of fundamental importance.

 Although naively the photon mass is expected to be directly restricted from the uniformity-lengths of the astrophysical magnetic fields of the galaxy and other sources, this was shown not to be the case~\cite{ADG}. The reason is that a non-zero photon mass fundamentally changes the topological structure of the vacuum leading to the 
  formation of Nielsen-Olesen~\cite{Nielsen} type magnetic vortex lines.
    An elementary photonic vortex line (or a string) represents a magnetic flux tube of the width given by 
   the photon's  Compton wavelength, $l_{\gamma} \equiv \hbar /m_{\gamma}c$ (where $c \equiv$ speed of light), with a much 
  thinner Higgs core in the middle.   
 As shown in~\cite{ADG}, the vortexes ensure that a strong magnetic field can be effectively Maxwellian
 over distances  $\gg l_{\gamma}$.  The reason is that such an uniform Maxwellian field is composed out of large number  of vortexes with overlapping magnetic cores. 

  This finding has the following two effects. On the one hand, it makes the uniformity of the galactic  magnetic field  fully compatible with values of the photon mass relevant  for much shorter scales of astrophysical interest.  
In particular, the range of $l_{\gamma}$ 
permitted by 
the scenario of~\cite{ADG}, fits the  
sizes of supermassive black holes. 

On the other hand, the vortex structure opens up a new mechanism for generating a strong magnetic field. The maximal strength is determined by the capacity of a given object to hold together a large number of overlapping vortexes. 
We point out that important representatives of this 
   category are black holes.  
    Due to its gravitational pull, a black hole can trap a large number of overlapping flux tubes, thereby, enhancing
   its magnetic field.  We shall refer to this mechanism 
  as the ``vortextrap magnetization" (VTM).

 The VTM mechanism can be 
   efficient to the extent that in the vicinity of the event horizon the magnetic field might reach the maximal sustainable value, 
\begin{equation} \label{BBound}
    B \sim \frac{c^4}{MG^{3/2}}\,, 
\end{equation}
 where  $M$ is the black hole mass 
 and $G$ is the Newton's constant.
 
 Notice that the above expression represents  an absolute upper bound  
 on the magnetic field of a black hole, regardless of its origin, 
 which can be imposed from the following energetic 
considerations. Indeed,
we arrive to this value by equipartitioning the gravitational self-energy of a black hole,  $E_{\rm gr} \sim Mc^2$, and the energy of the magnetic field, $E_{\rm mag} \sim B^2 R^3$, where 
$R \sim MG/c^2$ is the black hole's gravitational radius. 
Equating $E_{\rm mag} \sim E_{\rm gr} $,  gives (\ref{BBound}). 
 This bound is universal, independently of the value of the photon mass.  However, the existence of a non-zero photon mass 
    influences the situation in two ways. 
  
 First, in form of VTM it provides a concrete mechanism 
 for sustaining a state with a strong magnetic field due to the collective 
 magnetic flux of photon vortexes flowing through a black hole.    
  The strength of this collective field can reach the bound (\ref{BBound}).
 
 Secondly, the VTM mechanism suggests that the accumulation by a black hole of the magnetic field above a certain threshold value 
is expected due to the trapping of the photon vortexes encountered  
during the cosmological evolution. We estimate that the minimal strength of 
the magnetic field acquired in this way can be rather 
significant and competing with 
the values usually assumed to be generated via conventional mechanisms.

 The possession of a strong magnetic field by a black hole can result into  a number of important effects.
In particular, it is of direct relevance for various mechanisms
of particle acceleration. 
The famous examples are the so-called Fermi acceleration~\cite{fermi} with its several modifications~\cite{bell1,bell2,catanese} as well as 
the Blandford-Znajeck mechanism~\cite{bz, bz1}.

 Another scenario for which VTM has direct implications is 
  so-called magneto-centrifugal acceleration (MA)~\cite{orb,newast,rm}. 
  This mechanism is based on the 
assumption that  plasma particles are in frozen-in conditions, 
meaning that they follow the rotating magnetic field lines.  
 The role of MA for active galactic nuclei (AGN) has been studied in a series of papers~\cite{orb,newast,rm} and it was shown that by means of direct centrifugal acceleration, particles might achieve several TeV energies.

 Understanding of particle acceleration mechanisms is of particular interest in the light of very high-energy cosmic rays.  
It is believed that protons with energies $>10^{17}$ eV might be produced by AGN~\cite{kim}. Such protons, when interacting with each other can potentially produce the PeV energy neutrinos~\cite{murase} detected by the Ice Cube collaboration~\cite{neutrinos, Icecube1, Icecube2}.  Recently, the KM3NeT collaboration reported the detection of a $220\,$PeV neutrino (KM3-230213A)~\cite{KM3NeT:2025npi}, the most energetic neutrino observed to date. This detection further supports the existence of astrophysical accelerators capable of producing cosmic rays and neutrinos at extreme energies. For a comprehensive multi-messenger study of IceCube neutrinos, we refer the reader to~\cite{IcecubeR}. 
In summary, the study of the origin of ultra high energy protons opens a window also for exploration of astrophysical neutrinos.

  In the current work we also point out that a near-saturated value of the magnetic field can have important implications for black hole mergers, 
  since the substantial fraction of the magnetic energy 
  (comparable to a black hole's gravitational self-energy) is expected to be released in a high-intensity electromagnetic radiation as well as in secondary gravitational waves. 
  
   Finally,  we illustrate the dynamics of VTM 
 both analytically and numerically on a simple prototype model.
 In this model the role of a vortex-trapping  black hole is played by 
 a solitonic vacuum bubble, which was introduced as an effective 
 description of a black hole in the earlier papers~\cite{Saturon2020, S4,DvaliV1,DvaliV2} within the framework of a  so-called ``saturon" program.   
  The point is that a class of solitons (referred to as ``saturons" 
 in~\cite{Saturon2020}), while being much simpler to analyse, 
   at the same time shares the relevant key features 
  with black holes. Correspondingly, it represents an useful laboratory 
  for understanding the various aspects of the black hole dynamics.   
   The presented analysis 
  of VTM in this prototype system fully matches the expected dynamics 
  in black holes. 
  
  We must remark that, besides providing an useful lab for  understanding VTM in black holes, the highly-magnetized 
  vacuum bubbles can be interesting astrophysical objects in their own rights, as they can lead to the phenomena very similar to their black hole 
  relatives.  
 
   Before proceeding, we would like to clearly isolate the
   presented VTM mechanism  from  the alternative scenario of black hole magnetization 
    via vorticity introduced in~\cite{DvaliV1,DvaliV2}. 
   In this work it was conjectured that a highly-spinning black hole develops an internal gravitational vorticity which 
   prevents any further growth of the spin relative to the mass. 
   Among other things, this explains the extremality relation between the maximal  spin of a black hole and its mass (or entropy).  
  It was also argued that,  when interacting with a surrounding 
  plasma, the gravitational vortex can trap a magnetic flux and effectively become a magnetic vortex.
Various implications of this scenario such as particle acceleration or primordial magnetic fields require independent studies~\cite{DKO}. 
  
   Here we wish to comment that the effect of black hole magnetization via the internal  gravitational vorticity 
  introduced in~\cite{DvaliV1}
  is very different from VTM mechanism considered 
  in the present paper. First, the former effect requires a near-maximal spin of a black hole and postulates its connection with internal gravitational vorticity.  Secondly, this mechanism is unrelated 
  to the photon mass.   
 In contrast, the VTM scenario considered in the present work does not rely on the spin of a black hole and also makes no assumptions about possible existence of the gravitational vorticity. In fact, we make no assumptions about the internal microscopic structure of a black hole. 
  Instead, the sole assumption is a non-zero mass of the photon. 
  However, we must say that despite the fundamental
differences between the two scenarios, they can have 
some related implications, in particular for black hole mergers~\cite{SaturonV}, due to the shared  predictions of the magnetic vorticity.

The paper is organized as follows. 
In section~\ref{sec2:photonmass},  we discuss the essential features of magnetostatics 
in the presence of the photon mass and the role of vorticity.  
In section~\ref{sec3:bhmagnetization},  we  describe VTM mechanism and show how 
it can lead to saturation of the bound (\ref{BBound}) by a magnetic field of 
a supermassive black hole. 
In section~\ref{sec4:cosmologicalscenario}, we discuss a minimalistic VTM
scenario of generation of the magnetic field of a black hole by trapping of vortexes 
during the cosmological history. 
In section~\ref{sec6:multi}, the role of the saturated magnetic field is discussed in light of the multi-messenger physics of black hole 
merger. In section~\ref{sec7:multimess}, we illustrate the VTM effect on a prototype   
system. 
Finally, in section~\ref{sec8:conclusion} we summarize our results.

Visuals of the dynamics of the VTM mechanism can be found at the following \href{https://youtu.be/KnxQLbEHmuU}{URL}.

\section{Photon mass and magnetic field} \label{sec2:photonmass}

 In this section, we shall present some relevant information about the photon mass and fix the framework introduced in~\cite{ADG}. 
 We shall use the units $c=\hbar =1$.  In particular, in this units, 
 $l_{\gamma} = m_{\gamma}^{-1} $.

  A frequently referred upper bound on the photon mass, $m_{\gamma}\, \sim \,  10^{-27}$eV,    
  relies on the coherence length of the galactic magnetic field~\cite{Galaxy1,Galaxy2}.  
  The basic assumption in derivation of this bound is that the galactic magnetic field, $B_g$,  with  uniformity length $l_g$, would be accompanied  by the Proca energy density $\sim m_{\gamma}^2(B_g l_g)^2$, which is constrained by observations.  Notice that this Proca energy would also be constrained  by the experiments based on measurements with the toroidally magnetized pendulum in a magnetically shielded  vacuum~\cite{Lakes:1998mi, Luo:2003rz}, resulting in a milder  upper  bound $m_{\gamma} \sim 10^{-16}$eV.

    However, as it was shown in~\cite{ADG}, the above reasoning 
is not applicable for a very large portion of the theory parameter 
space, as the existence of the galactic Proca energy is highly  
sensitive to the microscopic origin of the photon mass. 
 Correspondingly, the uniformity-length 
of the magnetic field is not directly translatable into the bound on 
the photon mass.  Namely, the would-be Proca energy is cancelled  
by the winding of the longitudinal polarization of the photon,
resulting in formation of vortexes. As a result,  the system 
can support an uniform magnetic field on scales $l \gg m_{\gamma}^{-1}$.

The fundamental reason is that $m_{\gamma} \neq 0$ affects the 
would-be Maxwell theory in the following way: 
  \begin{itemize}
  \item The number of degrees of freedom 
changes discontinuously from $2$ to $3$, since the massive photon acquires a longitudinal polarization. 
  \item  The topological structure of the vacuum changes allowing for the existence of magnetic vortexes. 
 \item  Due to vorticity, while the electric field 
is screened at distances exceeding  $m_{\gamma}^{-1}$, the 
magnetic field is not. 
  
\end{itemize}

Thus, unlike the electric field, the 
magnetic one is not screened by the photon mass. 
Instead, the magnetic field is confined into the flux 
tubes that represent Nielsen-Olesen type vortex lines. 
The elementary flux tube has the width $m_{\gamma}^{-1}$
and can extend to infinity. 
 The multiple overlapping tubes  can create a magnetic field 
with an arbitrarily large coherence length. 
  In order to understand the story let us closely follow~\cite{ADG}.  
   
   We start with discussing the physical meaning 
   of the photon mass.        
  The Standard Model of particle physics allows for an unique  consistent extension of the Higgs effect in which the photon, which we shall denote by $A_{\mu}$, gets a small mass.  
This requires an additional
Higgs field in form of a complex scalar $\Phi$ 
with non-zero electric charge \footnote{At the level of the full 
Standard Model, the electric charge of $\Phi$ originates from its 
hypercharge. There is no need to enter in this construction, 
since the low energy effective theory relevant for 
astrophysical scales is unique (see below).}. 
It is convenient to write the field $\Phi$ in terms of its modulus, $\rho(x)$,  and the phase, $\varphi(x)$, as 
$\Phi(x) = \rho(x) e^{i\varphi(x)}$. 
  
  The unique effective low-energy theory describing the massive photon is given by the standard Higgs Lagrangian, 
     \begin{equation} \label{HiggsL}
  \mathcal{L}_{\rm Higgs} \, = \, |D_{\mu}\Phi|^2 - \frac{1}{4}\lambda^2\left(|\Phi|^2 - v^2\right)^2
  - \frac{1}{2} F_{\mu\nu} F^{\mu\nu}
   \end{equation} 
  where $D_{\mu} \equiv \partial_{\mu} - ig A_{\mu}$ is the covariant derivative and 
  $F_{\mu\nu} \equiv \partial_{\mu}A_{\nu} - \partial_{\nu}A_{\mu}$ is the Maxwellian field strength (non-standard normalization is for  later convenience).
  Here  $g$, $\lambda$ are dimensionless coupling constants, whereas the parameter $v$ has a dimensionality of mass.
  Notice that $g$ is a product of the ordinary 
  electromagnetic gauge coupling and the charge of $\Phi$ 
  which must be minuscule due to phenomenological reasons~\cite{ADG}.   
  
  In terms of the phase and the modulus fields, the Lagrangian 
  can be rewritten as, 
     \begin{equation} \label{HiggsL1}
  \mathcal{L}_{\rm Higgs} \, = \, (\partial_{\mu}\rho)^2 + (g\rho)^2 
  \left( A_{\mu} - \frac{1}{g} \partial_{\mu}\varphi\right)^2
 -   \frac{1}{4}\lambda^2(\rho^2 - v^2)^2
  - \frac{1}{2} F_{\mu\nu} F^{\mu\nu}\,.
   \end{equation} 
 Obviously,  the theory exhibits the following $U(1)_{\rm EM}$ gauge redundancy, 
     \begin{equation} \label{GaugeS}
  A_{\mu}  \, \rightarrow  \,   A_{\mu} \, + \, \frac{1}{g} \partial_{\mu} \alpha\,,  
  ~ \, \varphi \rightarrow \varphi  + \alpha  
   \end{equation}   
 where $\alpha(x)$ is an arbitrary function.

   As it is well-known, the vacuum of the theory is described
 by a constant vacuum expectation value (VEV) of the modulus $\langle \rho \rangle =v$. 
   Correspondingly, the $U(1)_{\rm EM}$-theory is in the Higgs phase.  The would-be Goldstone mode $\varphi$ becomes
   a longitudinal polarization of the vector boson and 
   the two form a massive photon field, with the mass 
   $m_{\gamma} = g\,v$.  The fluctuation of the modulus, 
   $\rho(x) = v + \frac{1}{\sqrt{2}} h(x)$, represents a  massive Higgs boson, with the mass $m_h = \lambda\, v$.

   For the phenomenological reasons of non-observation 
   of the $h(x)$-boson, the parameter $g$ must be very small.   
   The parameter $\lambda$ is less constrained  and can be taken all the way up to its perturbative bound 
   $\lambda \sim 1$. 

At energy scales below $m_h$, the effective theory of the massive photon reduces to the Proca theory,  
     \begin{equation} \label{ProcaL}
  \mathcal{L}_{\rm Proca} \, = \,  \frac{1}{2} m_{\gamma}^2 \tilde{A}_{\mu}\tilde{A}^{\mu}   
  - \frac{1}{4} F_{\mu\nu} F^{\mu\nu} \, 
   \end{equation}   
  where $\tilde{A}_{\mu} \equiv \sqrt{2}\, (A_{\mu} - \frac{1}{g} \partial_{\mu}\varphi)$  is the massive Proca field. 
  
  As it is clear, the longitudinal polarization of the Proca field
  is originating from the phase of the Higgs field. 
  Notice that the gauge redundancy (\ref{GaugeS}) of the 
  Higgs Lagrangian  is fully 
  inherited by the Proca theory.   
  This clarifies a frequent misconception 
  about the breaking of the gauge  symmetry by the photon mass. 
 In reality, the gauge redundancy remains intact but becomes 
 hidden due to the gauge-invariance of $\tilde{A}_{\mu}$. 
 In Proca theory 
  the longitudinal polarization of the photon, 
  $\varphi$, plays the role of the St\"uckelberg field 
  which maintains the gauge redundancy of the ``parent" Higgs theory
  (see, e.g.,~\cite{Dvali2005}).

 The effective Proca Lagrangian (\ref{ProcaL})  is sufficient for 
 accounting for perturbative physics of a massive photon 
 at distances larger than the Compton wavelength of the Higgs 
 field, $m_h^{-1}$.  
  However, as pointed out in~\cite{ADG}, it is not capable of accounting for some important non-perturbative physics. 
    In particular, in order to understand  how the configurations  with the magnetic field are affected by a non-zero photon mass $m_{\gamma}$, 
    we must take into account the Nielsen-Olesen vortex lines. 
  For achieving this, one has to incorporate the dynamics of the Higgs field
  by upgrading the Proca theory (\ref{ProcaL}) 
  into the Higgs theory (\ref{HiggsL1}). 
   This non-perturbative physics removes restrictions from 
   the coherence length of the magnetic field, and 
   correspondingly lifts the naive bound on $m_{\gamma}$.

    The Nielsen-Olesen vortex lines represent the magnetic 
    flux tubes analogous to Abrikosov vortexes in 
    superconductors.  The isolated 
    vortex solution with a winding number $n$ 
    in cylindrical coordinates 
    $r, \theta, z$ corresponds to the following ansatz, 
    \begin{equation} \label{ansatz}
    \rho(r)\,, ~~~\varphi = n\,\theta\,, ~~~ {\rm and} ~~~  A_{\theta}(r)\,,
    \end{equation}
    with all other  components zero. 
     The asymptotic values of the Higgs field are 
     $\rho(0)=0$, and $\rho(\infty) = v$, where 
     the asymptotic value $v$ is approached
     for $r \, > \,  m_h^{-1}$. 
     
     At the same time, for $r >  m_{\gamma}^{-1}$, the gauge field $A_{\theta}(r)$  approaches a locally-pure-gauge form, 
     \begin{equation} \label{pgauge}
         A_{\theta}(r) \, = \, \frac{n}{g\,r} \, . 
    \end{equation}   
     This results into a magnetic field $B_z = d_r A_{\theta}(r)$. 
     The magnetic flux flowing in $z$-direction 
     can be found from Stokes' theorem and is given by,  
       \begin{equation} \label{flux}
    {\rm Flux}=\frac{n}{g}  \,. 
  \end{equation}   
  Correspondingly, the magnetic flux is quantized in units of $1/g$.  
 
 Thus, the vortex has two cores: the Higgs core 
     of radius $m_h^{-1}$ and the magnetic core 
     of radius $m_{\gamma}^{-1}$. 
   Correspondingly,  the energy of a vortex 
    can be split into the Higgs and magnetic energies. 
    Up to log-factors, both energies per unit length 
    of the vortex line are of order $v^2$. This sets the 
    string tension, defined as the energy per unit length of the vortex line, 
    to the value
    $\mu \sim v^2$.

  The average value of the magnetic field is, 
     \begin{equation} \label{Bofnv}
        B \sim  \frac{n}{g} \, m_{\gamma}^2 \,.
        \end{equation}
  This formula for the magnetic field is also valid for 
  $n$ elementary vortexes with overlapping magnetic cores.   
       
   While the Nielsen-Olesen vortex configuration is obtained as an explicit solution of the equations of motion, 
   the above behaviour can be understood 
  by balancing the various entries in the two-dimensional energy density functional, 
 \begin{equation} \label{Etr}
  \mu = 2\pi \int_0^{\infty} { d}r\, r \left[ ({ d}_r\rho)^2
  + \rho^2(gA_{\theta} - n/r)^2 + \frac{1}{4}\lambda^2(\rho^2 -v^2)^2
  + B_z^2\right]\,.   
  \end{equation}  
  It is clear that the Higgs energy vanishes 
  outside of the Higgs core where we have 
   $\rho=v=\,$const.  Similarly, the magnetic energy vanishes 
   outside of the magnetic core 
   where $A_{\theta}$ is given by 
  (\ref{pgauge}).

     The above makes it clear why a non-zero photon mass 
     is not in conflict with the uniformity 
     of the magnetic field over the scales arbitrarily larger than 
     $m_{\gamma}^{-1}$. 
     The reason is that the would-be energy of the vector potential  $A_{\theta}$ of a magnetic  field  in the second term of (\ref{Etr}) is minimized 
     by the winding of the Goldstone phase $\varphi$. 
  That is, instead of restraining the magnetic field, 
  the phase adjusts to it by winding and thereby creating the vortex configurations.       
    As a result,  an uniform magnetic field is sustained 
    by a large number of overlapping vortexes~\cite{ADG}. 
     In particular, a homogeneous distribution of the vortexes 
  with $n$ overlapping magnetic cores, produce the magnetic field of the strength (\ref{Bofnv}).

 Following~\cite{ADG}, we can understand the essence of the effect 
 as the dynamical lowering of a would-be Proca energy of a constant magnetic field $B_z =  B$ via the vortex formation. 
 In the absence of vortexes, i.e. in the Higgs vacuum with
 $\rho = v, \varphi = const$,  the corresponding vector potential 
 $A_{\theta} = B\,r$ would result into a divergent Proca energy. 
 Namely, within a circle of radius $R$, the 
 Proca energy would scale as  
  \begin{equation} \label{Epr}
  E_{\rm Pr} = 2\pi \int_0^{R} {d}r\,r ~ 
  \left(g\,v\right)^2\left(A_{\theta} - \frac{1}{gr}d_{\theta} \varphi\right)^2  =  \frac{\pi}{2} \left(gvB\right)^2R^4 \,.
  \end{equation}  

  However, this energy is cancelled by the winding of the 
  phase $\varphi$ with the  average relation,  
  $d_{\theta} \varphi = gBr^2$. 
  
  This implies the existence of uniform distribution of zeros of the Higgs field. They represent the Higgs cores of the Nielsen-Olesen vortexes. 
  Around each zero the phase winds by $2\pi$. The winding number 
  around a closed circle of radius $r$ is given by, 
    \begin{equation} \label{Nr}
    N(r)  = \frac{1}{2\pi} \int  { d}_{\theta} \varphi \,.
  \end{equation}  
   Obviously, since the winding number has to be an integer, the cancellation cannot be exact and the residual Proca energy density 
   is $ \sim  gBv^2$.  If this energy exceeds the Higgs energy, with density $E_{\rm Higgs} \sim \lambda^2 v^4$,  it is energetically preferable
 for the system  to make the Higgs VEV zero everywhere. 
 That is, the magnetic field is sufficiently strong for pushing  
 the Higgs VEV to the top of the Mexican hat potential. 
 In other words, in such a regime the magnetic field un-Higgses the theory.

 This fixes the 
   critical value of the magnetic field, 
     \begin{equation} \label{Bcrit}
    B_c \, = \, \frac{1}{g}\lambda^2v^2 \,. 
  \end{equation}  
   For $B > B_c$ the vortexes become so densely populated that their 
   Higgs cores start to overlap.  This forms an uniform restored symmetry phase. 
   
     For $B < B_c$ there are different regimes of interest that depend on the size of the system.       
     For  $R > \frac{1}{\sqrt{gB}}$, the vortexes are energetically 
     favourable.  However two sub-regimes are possible.  
    For $R > \frac{\lambda v}{gB}$, the system classically 
    creates vortexes out of the vacuum in order to cancel the Proca energy.    
   In the opposite case   $R <  \frac{\lambda v}{gB}$, the vortexes are 
   still energetically favourable but cannot be created classically.  
   Various regions of the parameter space have been explored in numerical simulations~\cite{MaxJuan}.

   Both regimes are of astrophysical 
   importance. In particular, the photon vortexes, 
   that currently make up a particular astrophysical magnetic field, could have been pre-existing in the Universe as a result of an earlier phase transition. 
   Alternatively, they can be created by the magnetic field
 of an external source such as a star or a galaxy. 

For  $v \gg m_{\gamma}$,  the magnetic cores 
   of the vortexes can overlap even if the Higgs cores are 
   far apart. In general, the number $n_v$ of separated Higgs vortexes with overlapping magnetic cores is bounded by 
   \begin{equation} \label{NV}  
   n_v  \sim  m_h^2/m_{\gamma}^2\,. 
   \end{equation} 
     For a higher density of the vortexes the  Higgs cores start to overlap and the system can enter the un-Higgsed phase 
     as discussed above. 
 For $n_v$ magnetically-overlapping vortexes, the energy cost per area of the magnetic core  ($\sim m_{\gamma}^{-2}$) is $\mu \sim n_v^2 v^2$.
      
   Now, the point of~\cite{ADG} is that for $n_v \gg 1$ the vortexes produce a magnetic field which is practically indistinguishable from the Maxwellian case. This lifts the naive uniformity 
   bound on the photon mass.  According to this 
   picture, a magnetic field of certain coherence length  
   $l \gg m_{\gamma}$, in reality represents a ``forest" of  uniformly distributed photonic vortexes with overlapping  magnetic cores.

   In summary~\cite{ADG}, the vortex dynamics originating from 
   the photon mass has a dual implication.   
  First,  the bounds on the photon mass from
  astrophysical magnetic fields are lifted, leaving us with milder upper bounds such as the ones coming 
  from the direct measurement of the Coulomb law~\cite{CoulombB}
  $m_{\gamma} <  10^{-14}$eV, or the dispersion of the distant star light 
  $ m_{\gamma} < 10^{-17}$eV.  
   The corresponding Compton wavelengths of photon are within the scales of interest of supermassive black holes.    
  
  Secondly, when overlapping in large numbers, the photon vortexes can produce a very intense magnetic field.  As we shall discuss next, this overlap, supported by a black hole, can result into a magnetic field strength close to saturation (\ref{BBound}).

     Before proceeding, let us briefly comment on the gravitational effects of the photon vortexes. 
 An isolated  straight and infinite vortex line outside of its core creates a metric of a cosmic string~\cite{Vilenkin:1981zs}: 
     \begin{equation} \label{StringM}
    ds^2 \, =  dt^2\, - dz^2 - dr^2 - (1- 4G\mu)^2r^2d\theta^2  \, ,
  \end{equation}  
 where $\mu \sim v^2$ is the tension. 
 This metric is locally-flat but conical with the angular deficit  
 $\Delta \theta = 8\pi G\mu$. If $n$ parallel vortexes are put on top of each other, 
 the deficit will be controlled according to the collective tension.

   For $m_{\gamma}^{-1}$  less than the black hole radius, 
   the entire vortex line can go through a black hole. 
   The configuration is expected to be very similar to a black hole pierced by a cosmic string. 
In thin string approximation, the gravitational metric of such a system  was originally given by Aryal, Ford and Vilenkin~\cite{BHstring}. 
Their metric effectively 
superimposes the string metric on a Schwarzschild background 
renormalizing the black hole's internal energy 
due to the string deficit angle. 
 
     For us, a particularly relevant case  
    is a black hole with an abelian 
    Nielsen-Olesen vortex studied 
by Achucarro, Gregory and Kuijken~\cite{BHstringV}. This 
study represents a field theoretic resolution of Aryal-Ford-Vilenkin metric. 
 The black holes pierced by cosmic strings have been further discussed in the literature in various contexts (see e.g.,~\cite{Capture1, Capture2, Capture3, Capture4, Capture5, KerrString1, KerrString2}). 

 We must remark that there exists a very general consistency argument~\cite{Dvali:2008rm} according to which a stationary 
 groundstate configuration of a black hole and of an infinite thin string 
is the one in which the string pierces
 the black hole. That is, if the two are separated initially, the system will evolve towards the configuration of the black hole pierced by the cosmic string.  
 This argument is based on a version of a black hole no-hair theorem 
 according to which a (semi)classical black hole must couple democratically to all the particle species.  In other words, 
 a stationary classical black hole must posses no ``species hair"~\cite{Dvali:2008fd, Dvali:2007hz, Dvali:2007wp}. All astrophysical black holes  fall under the above constraint.

  For  $m_{\gamma}^{-1} > R$ the black hole can be pierced 
  by the Higgs core, which in the regimes of our interest 
 is much smaller than $R$, whereas the magnetic core can spread 
 outside of a black hole.  
 We shall be working in a regime 
   in which the would-be deficit angle is well bellow $2\pi$. 
    Correspondingly, the collective energy of vortexes shall not exceed 
    the mass of a black hole. 
    
     Notice that in cases in which the photonic strings captured by a black hole can be substantially long, they can result into a 
 observable double-image lensing effects characteristic to 
  a heavy cosmic strings.   Remarkably, even for a 
  sub-critical magnetic field, e.g., $B \sim  10^{-5}B_c$, 
  the collective lensing effect would be at the current observational bound~\cite{Chernoff:2007pd}.

  \section{Black hole magnetization by trapping  of photon vortexes} \label{sec3:bhmagnetization}

  In understanding the role of the photon mass for endowing the  black holes with strong magnetic fields, we split the discussion in 
  two separate questions.  The first step is to understand the 
  VTM mechanism of achieving the highly-magnetized black hole states in case of a massive photon.  The second step is the outline of possible 
   cosmological scenarios leading to the formation
  of such states throughout  the Universe's history.  
   In the present section we shall address the first question.

      According to~\cite{ADG}, if photon has a Compton wavelength 
 shorter than the uniformity length of a given astrophysical magnetic field, the effectively-Maxwellian field  
 is sustained by a large number of vortexes with overlapping 
    magnetic cores.   That is, the 
    Universe is populated by a high density of magnetic vortexes 
   of Nielsen-Olesen type. The density of the vortexes varies in 
   accordance with an ambient magnetic field. 
 Correspondingly, a magnetic capacity of a given object
 is determined by the ability to accumulate and sustain a 
 large number of overlapping photon vortexes. 
 
 In case of a black hole, this capacity is determined by
 its gravitational power to trap and hold vortexes passing through its interior~\cite{BHstring, BHstringV}.
  Interestingly, there is a curious coincidence that the sizes of
 supermassive black holes, are around or above the current experimental 
 bound on the photon's  Compton wavelength advocated 
 in~\cite{ADG}.  This allows for the phenomenologically-viable scenario of saturation of their magnetic fields by photon's vortexes.  

 That is, our basic point is that a non-zero photon mass leads to 
     the existence of black holes pierced by a large number 
     of overlapping ``fat" magnetic flux tubes.  
  As we shall estimate, this allows a black hole 
  to posses a magnetic field near saturation~(\ref{BBound}). 
We shall not make any assumptions about the microscopic properties of the black hole interior and 
shall solely rely on known semi-classical features of 
black hole's gravity. 
   
    We shall consider a situation when overlapping 
    magnetic flux tubes pierce through a black hole of mass $M$ and radius $R = 2M/M_{\rm P}^2$  (where $M_{\rm P} \equiv 1/\sqrt{G}$ is the Planck mass). 
    In order  to understand the value of the attained magnetic 
    field, we parameterize  the gauge coupling of 
 the  Higgs as $g = k\, m_{\gamma}/M_{\rm P}$, where
$k$ is a parameter, which can be written as, 
\begin{equation}\label{NandV}
 k = \frac{M_{\rm P}}{v} \,.
\end{equation} 
Obviously, in all the cases of interest $k$ is an extremely large number. 

Let us explore a black hole of radius $R$ pierced by $n_v$ vortexes with overlapping magnetic cores. We wish to estimate the values of  $n_v$ and $k$ for which the resulting magnetic field 
 saturates the bound (\ref{BBound}). In terms 
 of $M_{\rm P}$ and $R$, this saturated value 
 can be written as, 
  \begin{equation} \label{Bmax}
  B_{\rm max} \sim \frac{M_{\rm P}}{R}\,.  
    \end{equation}
   Equating the magnetic field of $n_v$ overlapping vortexes 
   to $B_{\rm max}$,  we get the condition for saturation
   \begin{equation} \label{satC1}
  B \sim n_v \frac{m_{\gamma}^2}{g}  
  \sim  \frac{n_v}{k} (m_{\gamma}R) B_{\rm max}\,, 
   \end{equation}   
   where we have used (\ref{NandV}) and (\ref{Bmax}). 
  
  Thus, the maximal magnetic field is reached for 
   \begin{equation} \label{SatC2}
   \frac{n_v}{k} (m_{\gamma}R) \sim 1\,. 
   \end{equation}      
  Taking into account (\ref{NV}) and 
  (\ref{NandV}),  this can be translated as the   
condition on the parameters of the theory, 
 \begin{equation} \label{NVNG}  
    v \sim \frac{1}{\lambda} \left (\frac{gM_{\rm P}}{R}
    \right)^{\frac{1}{2}} \,. 
   \end{equation} 
 For illustrative purposes, we can make further simplified assumptions 
 on the parameters. For example, taking $\lambda \sim 1$
 and $R \sim m_{\gamma}^{-1}$, 
 the equation (\ref{NVNG}) becomes,
   \begin{equation} \label{NVN}  
 v \sim \left (\frac{M_{\rm P}}{R^2}\right)^{\frac{1}{3}} \,. 
   \end{equation} 
    
  Let us explore some numerical examples.   
 We first take $m_{\gamma} \sim 1/R  \sim  10^{-15}$eV which is around the experimental bound on the photon mass generated via the Higgs mechanism~\cite{ADG}.  This Compton wavelength is of the order of the gravitational radius of a million solar mass black hole
 ($R \sim 10^{11}$cm).   From (\ref{NVN}) we get 
$v \sim $eV.  Correspondingly, $k \sim n_v \sim 10^{28}$.  

 For, $m_{\gamma} \sim  1/R \sim  10^{-18}$eV, 
 (a radius of a billion solar mass black hole)  we get 
 $v \sim 10^{-3}$eV.  Correspondingly the 
 maximal magnetic field is reached for 
 $n_v \sim k \sim 10^{31}$. 
  
    From the above, it is evident that the
    magnetic flux tubes 
    in form of Abrikosov-Nielsen-Olesen vortexes
    originating from the photon Higgs mechanism~\cite{ADG}  can 
    support the states of astrophysical black holes close to 
   magnetic  saturation (\ref{BBound}), or equivalently (\ref{Bmax}). 
   
    As already said, the configuration in which the black hole is pierced 
     by photonic flux tubes can be viewed as an extreme case 
     of  a black hole pierced by an abelian 
    Nielsen-Olesen vortex studied 
by Achucarro, Gregory and Kuijken~\cite{BHstringV}. 
    The difference in the present 
    case is that the gauge core of the string in question is  unusually ``fat" and 
    carries an ordinary magnetic flux. The 
    combined flux can generate the magnetic field close 
   to saturation (\ref{Bmax}). Of course, for the extreme 
values of the magnetic field, the back-reaction on the metric becomes significant. However, for order of magnitude 
estimates the extrapolation of 
qualitative features of known vortex-black-hole metrics~\cite{BHstringV}
is expected to be valid. 
  
      \section{A cosmological scenario of black hole magnetization via VTM} \label{sec4:cosmologicalscenario}

  We have seen that the vorticity supported by the photon mass
 allows for endowing a supermassive black hole
 with a strong magnetic field via the VTM mechanism. 
   The question how such magnetized black hole states can be 
  actualized during the cosmological evolution requires a separate investigation. 
  Here we shall discuss a basic  
  scenario which must be operative 
 regardless of more involved possibilities. 
  Its essence is that a black hole is 
 expected to trap and accumulate a certain minimal amount of the 
 magnetic flux in form of vortexes collected over its life-time.

 Notice that in general all black holes with sizes exceeding 
 the Higgs core of the vortex will  be subjected to 
 VTM to certain extents.  Within phenomenologically-consistent 
 values of the Higgs mass, this can cover essentially all astrophysical black holes. 

 Indeed, according to~\cite{ADG}, the magnetic field of 
 the galaxy is composed out of high density of overlapping vortexes. 
   These vortexes then are expected to be captured by a black hole 
  upon an encounter. This leads to the enhancement of black hole's magnetic field via VTM. 
  
A configuration of a black hole pierced 
by a string~\cite{BHstring, BHstringV} is expected to be a probable  outcome of an encounter between the two.   
The fact that a thin string trapped by a black hole must be a groundstate 
of the system is also supported by 
the general ``no-species-hair" arguments~\cite{Dvali:2008rm}.
 Whether the kinematics of the encounter allows for relaxation  
 to such a groundstate, is a dynamical question.  
   Some dynamical aspects of capture of cosmic strings by a black hole have been 
   discussed previously in the literature~\cite{Capture1,Capture2, Capture3,Capture4,Capture5, Xing:2020ecz, Deng:2023cwh}. 
   The difference in our case is the hierarchy of sizes between 
   the magnetic and the Higgs cores. 
    While a magnetic core can be larger or smaller relative to 
the size of a black hole, the Higgs core is always taken to be microscopically smaller. 
   We shall assume that upon 
   their encounter the probability  of trapping  a vortex by a black hole is order one.  Basically, we assume that the capture cross section is geometric $\sim R^2$.

 Thus, the outline of the scenario is the following. 
   First, we adopt the picture of~\cite{ADG}, according to which the magnetic field of the galaxy, $B_{\rm g}$,  is composed out of a 
   large number of uniformly 
   distributed  magnetically-overlapping photonic vortexes.
  The origin of the galactic magnetic field  shall not be our concern.
   We shall simply take its existence as a fact.  The presence of vortexes 
   is not an additional assumption, since, as shown in~\cite{ADG}, 
   with the magnetic field in place, its vortex composition is a must. 
   
     The number of overlapping photonic vortexes reproducing this field is 
    \begin{equation} \label{ngalaxy} 
    n_g \sim  g \frac{B_{\rm g}}{m_{\gamma}^2} \,. 
    \end{equation} 
   A black hole of radius $R$ moving 
   through  this vortex forest, collects them upon the encounter. 

   Here we need to differentiate among the black holes 
   of solar and intermediate masses, which can travel through 
   the galaxy, and the supermassive ones that are ``stationed" near the center.  However, in both cases we assume that the encounter 
  between a black hole and the vortex network shall take place due to the relative motion.   That is, even a stationary black hole 
  is expected to be subjected to a ``wind" of the vortex lines
  due to the cosmological evolution of the network.

   Assuming that the probability of capture is order one, 
   after traveling a relative distance $l \gg R$, the number of collected vortexes by 
   a black hole is, 
    \begin{equation} \label{ncollect} 
    n_c \sim  (lR m_\gamma^2) n_g \sim 
    g (l R) B_{\rm g}\,, 
    \end{equation} 
where the factor $(lR m_\gamma^2)$ is the  transverse 
area swept by a black hole in 
units of the cross section of vortex's magnetic core.

On the other hand, the required number of vortexes piercing the black 
hole that would produce the saturated magnetic field $B_{\rm max}$
 (\ref{Bmax}) is, 
   \begin{equation} \label{nsatur} 
    n_v \sim g   \frac{M_{\rm P}}{R \,m_{\gamma}^2} \,. 
    \end{equation} 
 This gives, 
  \begin{equation} \label{ncnv1} 
    \frac{n_c}{n_v}  \sim  \frac{l B_{\rm g}}{M_{\rm P}} (R\,m_{\gamma})^2  \,. 
    \end{equation} 
 This fraction measures the accumulated magnetic field $B$ relative to 
 the maximal attainable value $B_{\rm max}$.  Therefore we can write, 
 \begin{equation} \label{ncnv2} 
    B  \sim  \frac{l B_{\rm g}}{M_{\rm P}} (R\,m_{\gamma})^2 \, B_{\rm max} \,. 
    \end{equation} 
  Now, the length of the traveled path by a black hole 
  relative to the vortex network throughout the Hubble time $t_{\rm H}$ can be estimated 
  as $l \sim 10^{-4} t_{\rm H}$. 
  Numerically this gives, 
    \begin{equation} \label{ncnv3} 
    \frac{n_c}{n_v}  \sim  10^{-6} (R\,m_{\gamma})^2 \,. 
    \end{equation} 
  Thus, from this mechanism, over a cosmological time a black hole 
  can  accumulate the magnetic field of the 
  following value,  
    \begin{equation} \label{BBmax} 
    B  \sim  10^{-6} (R\,m_{\gamma})^2 \,  B_{\rm max}\,. 
    \end{equation} 
We must reiterate that the value given above represents a minimal estimate of the magnetic field expected to accumulate around a black hole over its lifetime due to its exposure to a dense forest of photonic vortices. In our analysis, we assume that the black hole propagates through a region with an uniform magnetic field throughout its cosmological evolution. Namely, we assume that the black hole travels 
within the magnetic field of the galactic coherence length,
over time $t_{\rm H}$. 

Obviously, validity of this assumption depends on the object's precise trajectory as well as the cosmological evolution of the magnetic field.
In particular, the black hole may enter regions containing vortices with opposite winding numbers compared to those it has already trapped. This would lead to the capture of oppositely wound vortices, which could then annihilate with the pre-existing trapped magnetic fields. The detailed characterization of the imprints of this dynamical process is left for the future work.

Additionally, astrophysical black holes are always surrounded by plasma, which can further enhance the magnetic field. More sophisticated and model-dependent mechanisms could certainly be considered, but such an analysis is beyond the scope of this work.

 In this respect we wish to comment on an important difference between the  Higgs/Proca type fundamental mass of the photon and its effective ``mass" acquired from the  interaction with the astrophysical plasma. Surely, propagation through
an ionized medium generates 
an effective screening length which in certain 
aspects has an effect similar to a mass.  
 However, there are dramatic differences between the effective  mass generated in this way  and the 
 fundamental mass of a photon discussed in~\cite{ADG} and in the present work.  
 
    First, unlike the Higgs vacuum,  the ionized inter-stellar medium is not a superconductor.  Correspondingly, sustaining a stationary magnetic field requires an external source.  This applies to the magnetic flux 
   trapped in a vorticity generated  in such a medium. 
   Even if the flux-carrying vorticity is generated 
  in ionized plasma, maintaining it requires a constant external support. Without it, the vorticity 
 shall cease  to exist and the magnetic flux will spread and dissipate.   The story is fundamentally different in case of  the photon mass generated via the Higgs mechanism,  for  which the existence of vorticity is the vacuum effect. 
    
   Moreover, the  Higgs vacuum is different from an ordinary 
   superconducting medium, as there are no actual particles 
   in the vacuum condensate.  
   Due to this feature, the Higgs vacuum is Poincare-invariant and the photonic vortexes are  originating from its non-trivial topological structure. The Higgs core plays an absolutely crucial role  
 both in maintaining the magnetic flux tubes without any external help as well as in their trapping by a black hole.

\section{Implications of VTM for multi-messenger physics of merger} \label{sec6:multi}

We would like to briefly mention a rather straightforward potentially-observable implication of a  VTM-generated 
near-saturated magnetic field for black hole mergers.  During the merger it is 
expected that a substantial fraction of the magnetic energy will be released in form of the electromagnetic radiation. 
   Crudely, this process can be thought of as inter-commutation 
   of magnetic flux tubes piercing the black hole with subsequent 
   formation of closed flux lines and their decay into electromagnetic waves.

For a saturated value (\ref{BBound}), during the latest stage of the merger, the power of emitted radiation can be nearly 
Planckian, 
\begin{equation} \label{PowerPL}
L \sim \frac{M_{\rm P}c^2}{t_{\rm P}} \sim 10^{59} {\rm erg/s}\,, 
\end{equation}
where $t_{\rm P} \equiv \sqrt{\hbar\, G/c^5}$ is the Planck time and we temporarily reintroduced $\hbar$ and $c$. 
 This energy is expected to be released in electromagnetic radiation of wavelength 
 $\sim R$ of extremely high intensity.
  For example, for $M \sim 10^9 M_{\odot} $ the occupation number 
  of the emitted photons per wavelength $R \sim 10^{14}$cm is about 
  $N_{\gamma} \sim 10^{95}$ \footnote{ Notice that by no means the Planckian power 
 of the emission should be understood as if the emission process 
 represents a probe of quantum gravity in Planck scale regime. 
  The wavelengths of the emitted photons $\sim R$ are macroscopic
  and correspondingly their energies $E_{\gamma} \sim \hbar c/R$ are much less than the Planck energy $M_{\rm P} c^2$.
 The high power of radiation is a result of a very high amplitude,
 or equivalently, of a high occupation number of photons 
 $N_{\gamma}$. Correspondingly, the quantum gravitational effects are totally negligible and the emission process is well  
 within the validity of classical gravitational physics.
 For a general discussion clarifying the dependence of the strength of quantum gravitational effects on wavelengths of quanta versus field  intensity/occupation numbers, the reader is referred to~\cite{Nportrait}. }.   
  This electromagnetic radiation shall also source the secondary gravitational waves.  

   In the next section we illustrate the effect of the high-intesity  electromagnetic radiation during the merger
   in a numerical simulation performed on a simple prototype model
   which makes the essennce of the phenomenon very transparent. 
   This analysis confirms the qualitative expectations 
   given above.  The analysis of observational prospects is beyond the scope of the present paper, however,  the saturated magnetic field 
 is expected to make an order-one difference in the dynamics of 
 the merger.

 \section{Simulations of VTM in a simple prototype model}\label{sec7:multimess}

\subsection{Analytics} 
 
  In this section we  explicitly illustrate the VTM mechanism 
 on a simple prototype model that, while capturing  
 the key features of the mechanism, is easier to understand and solve  
 both analytically  and numerically.   
   This general approach  of simulating the dynamics of 
   black hole physics  in soliton systems 
   was used in several publications~\cite{Saturon2020,S1, S2, S3, S4,DvaliV1, DvaliV2} and relies on the finding
 that the fundamental features of black holes are shared 
 by a large class of solitonic objects, the so-called ``saturons"~\cite{Saturon2020}.  The defining feature of this class of objects is the maximal microstate degeneracy.  This correspondence allows to analyse many dynamical aspects  of black hole physics by using such solitons as laboratories.  
  
 In the present context we use this method for studying the VTM effect.  
 The idea is to replace 
 a black hole with a prototype soliton possessing a similar trapping power 
 of the electromagnetic vortex.   This role shall be assigned to a vacuum 
 bubble separating the phases with broken and  unbroken global symmetries~\cite{Saturon2020}  stabilized by the internal charge~\cite{S4,DvaliV1, DvaliV2}.
 
   In order to achieve this,  we introduce the additional scalar 
 field $\Sigma$ and add the following terms to the 
 Lagrangian (\ref{HiggsL})  
      \begin{equation} \label{Sigma}
  \mathcal{L}_{\Sigma} \, = \,  {\rm Tr}(\partial_{\mu} \Sigma)^2  - V(\Sigma)
    - \beta^2 ({\rm Tr}\,\Sigma^2)|\Phi|^2 \,,
   \end{equation} 
   where $V(\Sigma)$ is the  self-interaction scalar potential 
 of the $\Sigma$-field, whereas the last term describes the 
  cross coupling between $\Sigma$ and $\Phi$. Of course, it is assumed 
  that all terms are invariant under the symmetries acting on 
  $\Sigma$ and $\Phi$. 
 
    Both the internal quantum numbers of $\Sigma$ as well as the structure of $V(\Sigma)$ are chosen in such a way that 
  $\Sigma$ supports a continuous spectrum of the stable solitons~\cite{S4,DvaliV1, DvaliV2}.   
  The simplest class of such solutions emerges when 
  $\Sigma$  carries a charge $Q$ under some global symmetry and  $V(\Sigma)$ has two degenerate vacua with 
$\Sigma =0$ and   $\Sigma \neq 0$ respectively.  
  Obviously, these vacua correspond to the phases with unbroken and broken 
   $Q$-symmetry respectively.  In~\cite{Saturon2020,S4,DvaliV1, DvaliV2}, the field $\Sigma$ was chosen to transform
   as an adjoint representation of a global $SU(N)$-symmetry with 
  $Q$-charge identified with one of its generators. 
   We shall use this example  in numerical simulations below.  
   However, the VTM phenomenon is independent 
   on the representation content as long as the above-discussed general conditions are satisfied. 
   
   Following~\cite{S4, DvaliV1, DvaliV2}, 
   without any loss of generality, 
   we can parameterize the $\Sigma$-field as a complex 
   scalar $\Sigma (x)  = \sigma (x)  e^{i\chi(x)}$ with modulus 
   $\sigma$ and the phase $\chi(x)$ and reduce the system to its bare 
   essentials via the following effective Lagrangian, 
       \begin{equation} \label{SigmaEFF}
  \mathcal{L}_{\Sigma} \, = \,  (\partial_{\mu} \sigma)^2   + \sigma^2 
  (\partial_{\mu} \chi)^2   - \alpha\, \sigma^2(\sigma -f)^2
    - \beta^2 \sigma^2 \rho^2 \,, 
   \end{equation} 
   where $\alpha$ is the coupling and $f$ is a scale. 
   Obviously, the potential has two degenerate minima,  with the VEVs
   $\sigma = 0$ and $\sigma = f$ respectively. 
   In the vacuum with broken symmetry,  $\chi$ describes a massless Goldstone field,
  with the coupling (Goldstone decay constant) $1/f$,   whereas the perturbation of $\sigma$ has a mass $m_{\sigma}^2 = \alpha f^2$. 
    
  This system possesses  the vacuum bubbles of broken symmetry which have large internal Goldstone degeneracy~\cite{Saturon2020}
  \footnote{Remarkably, for the highest global symmetry permitted by unitarity (which for $SU(N)$ is saturated for $N \alpha =1$) 
  the bubble saturates the same universal bound 
on the microstate entropy as the black hole modulo the mapping 
$f \rightarrow M_{\rm P}$  \cite{Saturon2020, S4, DvaliV1}. This propety is the 
key to the black hole/saturon correspondence \cite{Saturon2020,
S1, S2, S3, S4, DvaliV1, DvaliV2}.
While this feature is unessential for the VTM mechanism, it further strengthens the 
motivation for using the above soliton as a test lab.}. 
 It has been further shown in~\cite{S4} that the vacuum bubble is  stabilized due to its internal rotation by the broken symmetry generator,  
 similarly to a non-topological soliton or a  $Q$-ball~\cite{QB1, QB2}. 
  In spherical coordinates, the field configuration has the form,
  $\Sigma = f(r) e^{i\omega t}$, where $f(0)  \neq 0$ and 
 $f(\infty) = 0$.   The parameter $\omega$ is arbitrary and determines the mass, $Q$-charge and the size of the bubble. 

  In order to match the regimes of interest motivated 
 by astrophysical  black holes, we shall work in a small back-reaction limit which is given by 
    \begin{equation} \label{backR}
     \beta v \ll m_{\sigma} \,~ {\rm and} \,~  v \ll f\,.
    \end{equation} 
 This choices guarantee that the influence from 
 the Higgs VEV on the bubble solution is negligible and the bubble is heavier than the energy of the elementary vortex line of the same size.    This is fully justified, since, translated to the gravitational counterpart, the condition (\ref{backR}) maps on the features that the Higgs vacuum does not affect the black hole and the vortex line is lighter than the black hole.

  Next, for clarity we shall start  with the approximation of a thin-wall and 
  then move to a generic case.  This implies that the bubble radius 
  $R$ is much larger than the wall thickness which is given by 
  the inverse mass of $\sigma$,  $ R \gg m_{\sigma}^{-1}$. 
   The mass $M$  and the radius $R$ of a stable thin-wall bubble are related as~\cite{S4}, 
        \begin{equation} \label{MandR}
    R \simeq \frac{M}{\pi \, f^2}  \frac{1}{m_{\sigma}R}  \,.
    \end{equation} 
  Notice that for a thick wall bubble, $m_{\sigma}R \sim 1$, 
  the above relation coincides with the analogous relation between the 
  black hole mass and the radius, with $1/f^2$ playing the role of the 
  Newton's constant (or equivalently,  $f$ assuming the role of $M_{\rm P}$). 
    This feature is no accident and lies in 
  deep universality of saturon solitons
  and black holes~\cite{Saturon2020,S1, S2, S3, S4, DvaliV1, DvaliV2}.  
    We shall not enter further in this correspondence but shall only use 
    the bubble as a laboratory for illustrating the VTM mechanism.

    A thin-wall bubble represents a sphere of vacuum 
    $\sigma = f$ embedded in the vacuum $\sigma = 0$.           
    We shall study the interaction between the bubble and 
    the photonic vortex-line produced by $\Phi$.    
    In the small back-reaction limit (\ref{backR}), the influence of the bubble on the Higgs field can be found by 
    minimizing the effective potential of $\rho$
   in the background of the bubble,  
         \begin{equation} \label{veffsigmarho}
  V_{\rm eff} (\rho)  = \, \beta^2 f(r)^2 \rho^2 \, + \, \frac{\lambda^2}{2} 
    (\rho^2 - v^2)^2 \,, 
   \end{equation}   
   where we have substituted $\sigma$ with the bubble 
   profile function $f(r)$. 
        
   The vacuum outside the bubble, $r > R$,  where $f =0$, 
   is identical to the previously-considered Higgs vacuum with the VEV 
   of $\rho$ given by,   $ \rho_{\rm out} = v$. Correspondingly, this vacuum 
   supports the Nielsen-Olesen vortex-lines as discussed above.
    The tension of the vortex line is $\mu_{\rm out} \sim v^2$. 
    In order to match the parameter regime of the previously-presented 
    astrophysical black holes, we shall assume that the thickness of the vortex's Higgs core, $\sim m_h^{-1}$,  is smaller than the bubble radius, 
  $R \gg m_h^{-1}$.

    Let us now discuss the vacuum in the bubble interior, $r < R$, 
     where $\sigma = f$.   For $\beta^2 > 0$, this lowers the VEV of the Higgs 
     as $ \rho_{\rm in} = v_{\rm in} = v \sqrt{1 - \frac{\beta^2 f^2}{\lambda^2 v^2}}$.     
    Of course, the inner vacuum still supports the vortex line but with a lower tension, 
    $\mu_{\rm in} \sim  v_{\rm in}^2$. 
    Correspondingly, the difference between the string tensions 
   in the bubble interior and the exterior is 
   \begin{equation} \label{mu}
  \Delta \mu \equiv  \mu_{\rm in} - \mu_{\rm out} \sim  - \frac{\beta^2 f^2}{\lambda^2}\,.
  \end{equation}   
   We can mentally think of this shift in the tension 
   as a scalar analog of the effect of the gravitational 
   ``red-shift".  
    
   Thus,  if a portion of the string of length $\Delta L$ enters 
   the bubble, the total potential energy of the system is lowered by $\Delta E(L) = - \Delta L  \Delta \mu$. 
    Therefore, there is an attractive force
    \begin{equation} \label{force} 
   F \, = \, \frac{\partial E}{ \partial L} \, = \,  - \Delta \mu , 
   \end{equation}   
    which pulls the vortex line inside the bubble.  Since inside the bubble the effective masses of the Higgs, $m_h = \lambda\, v_{\rm in}$, 
   and of the photon, $m_{\gamma} = g \,v_{\rm in}$,  are smaller than their vacuum counterparts, the portion of the vortex line absorbed by the bubble becomes thicker.  This fattening is exhibited both by the Higgs and the magnetic cores.  
     The ground state of the system of a bubble plus an infinite string is expected  to be an axially-symmetric  configuration with the vortex line piercing the bubble through the north and the south poles and with the inner portion being thicker than the outer one.  
    
   Notice that for  $\frac{\beta^2 f^2}{\lambda^2 v^2} >1$, 
   the Higgs VEV in the interior, 
  $v_{\rm in}$, is practically zero (in reality,  it is exponentially suppressed).  
  In this case the bubble wall becomes a boundary 
  between the Higgs and the Coulomb phases 
  of $U(1)_{\rm EM}$ theory.  If the vortex line crosses this boundary, 
  the magnetic flux, which in the Higgs vacuum is confined in a 
  tube,  in the Coulomb phase becomes spread.  The analytic and numeric studies of this phenomenon can be found  in~\cite{ErasureGJ}.

   Applying to the present case,  the vortex line piercing the vacuum 
   bubble will spread within the bubble interior.
   However,  the flux is conserved and continues to enter and exit the bubble   
   in form of the vortex line.       
  That is, passed the penetration point (say at the south pole), the flux spreads through the interior, reaching the maximum at the equatorial plane. Moving further up,  the flux is gathering back and is exiting the bubble through the north pole in form of the string. 
    
   Notice that the energetics of the bubble-string 
   interaction in the model (\ref{Sigma}) is determined by the Higgs core. 
   Correspondingly, the trapping tendency persists even if 
  the magnetic core of the vortex is much larger than the bubble radius. 
  
   All the above features can be smoothly extrapolated 
   away from thin-wall and zero back-reaction approximations.
    With this extrapolation the outcome of the  prototype system 
    fully matches the estimates for black holes obtained earlier in the  
   present paper.

    First, as explained, in the thick-wall regime  the  
    bubble parameters match the one of a black hole 
    with the identification $f \rightarrow M_{\rm P}$. In particular, the relation between the size and mass is, 
     \begin{equation} \label{MandR1}
     R \sim \frac{M}{f^2} \,.
    \end{equation}    
     Working in the regime 
    $v \ll f$ and $\lambda \sim 1$, we can see that the maximal value of the magnetic field  sustained by the bubble is  
    \begin{equation} \label{Bmaxf}
    B_{\rm max} \sim  \frac{f}{R} \,,
    \end{equation}  
   which with the substitution  $f \rightarrow M_{\rm P}$ 
   exactly matches the black hole case (\ref{Bmax}).   
     
    We shall now turn to numerical simulations of the above system.

      \subsection{Numerics}

In this subsection, we numerically illustrate the VTM mechanism using the simple prototype model~(\ref{Sigma}). Visuals of the dynamics we are about to discuss can be found at the following~\href{https://youtu.be/KnxQLbEHmuU}{URL} to which we refer the interested reader. 

We begin by examining how a bubble moving through a “forest” of magnetic flux tubes can reach the maximum magnetic field it can sustain, eventually leaving a portion of the excessive flux behind. We also show how the magnetic field evolves during the merger of two such saturated bubbles when pierced by a flux tube. As we will see, most of the magnetic energy is dissipated in the process.

Concretely, we consider a bubble configuration in a global $SU(N)$ theory, where the role of $\Sigma$ is played by an adjoint representation, i.e.\ a Hermitian traceless $N \times N$ matrix. Clearly, this choice does not affect the universal features of the VTM mechanism. 

The potential $V(\Sigma)$ in~(\ref{Sigma}) shall be taken as
~\cite{Saturon2020,S1, S2, S3, S4,DvaliV1, DvaliV2}:
\begin{equation}
\label{eq:potentialn}
  V(\Sigma) \;=\; \frac{\alpha}{2}\,\mathrm{Tr}\Bigl[f\,\Sigma \;-\;\Sigma^{2} \;+\;\frac{I}{N}\,\mathrm{Tr}\,\Sigma^{2}\Bigr]^{2}\,,
\end{equation}
where $I$ is the identity matrix, $\alpha$ is a dimensionless coupling, and $f$ denotes the characteristic scale. 

The model admits multiple degenerate vacua satisfying the matrix equality ($a,b$ are $SU(N)$ indexes):
\begin{equation}
\label{eq:vacua_conditionn}
 f\,\Sigma^{b}_{a} \;-\; \bigl(\Sigma^{2}\bigr)^{b}_{a} \;+\; \delta^{b}_{a}\,\frac{1}{N}\,\mathrm{Tr}\,\Sigma^{2} \;=\; 0\,,
\end{equation}
which exhibit different symmetry-breaking patterns.

We investigate vacuum-bubble configurations interpolating from the $SU(N)$-symmetric vacuum outside of the bubble to an $SU(N-1)\times U(1)$ vacuum inside the bubble. In the external region, the theory is in a gapped phase with mass squared $m^{2} = \alpha\,f^{2}$. Inside the bubble, the spontaneous symmetry breaking gives rise to $2(N-1)$ Goldstone species.  The bubble domain wall separating the two phases has a thickness $1/m$ and the tension $m^{3}/(6\alpha)$. 

Such configurations can be constructed via the ansatz:
\begin{equation}
\label{eq:rotationn}
  \Sigma \;=\; U^{\dagger}\,\Sigma_{\mathrm{rad}}\,U\,,
\end{equation}
where
\begin{equation}
\label{eq:phinew}
  \Sigma_{\mathrm{rad}} \;=\; \frac{\sigma(x)}{\sqrt{N(N-1)}}\,\mathrm{diag}\bigl[(N-1),\,-1,\,\dots,\,-1\bigr],
\end{equation}
and
\begin{equation}
\label{eq:Umatrix}
 U \;=\; \exp\Bigl[i\,\chi(x)\,\hat{T}/\sqrt{2}\Bigr].
\end{equation}
Here, $\hat{T}$ denotes one of the generators broken by the ansatz~(\ref{eq:phinew}), while $\chi(x)$ represents the corresponding Goldstone mode.
 Notice that by reducing the $\Sigma$-field to $\sigma(x)$ and 
 $\chi(x)$ components,  the system effectively reduces to  (\ref{SigmaEFF}).

We focus on solutions of the form 
\begin{equation}
\label{eq:bubblespherical}
 \sigma(x) = f(r)\,,\qquad \chi(x) \;=\; \omega\,t\,, 
\end{equation}
describing a spherical bubble. The function $f(r)$ satisfies $f(0)\simeq f$ and $f(\infty)=0$. The stability is ensured by the Goldstone charge $Q$. We solve for the bubble profile taking $\alpha=0.5$ and $\omega = 0.5\,m$. Moreover, for practical purposes, in the following we specialize for the case $N=4$.

\subsubsection{VTM Mechanism}

\begin{figure*}[!ht]
\centering
\includegraphics[width=1.\textwidth]{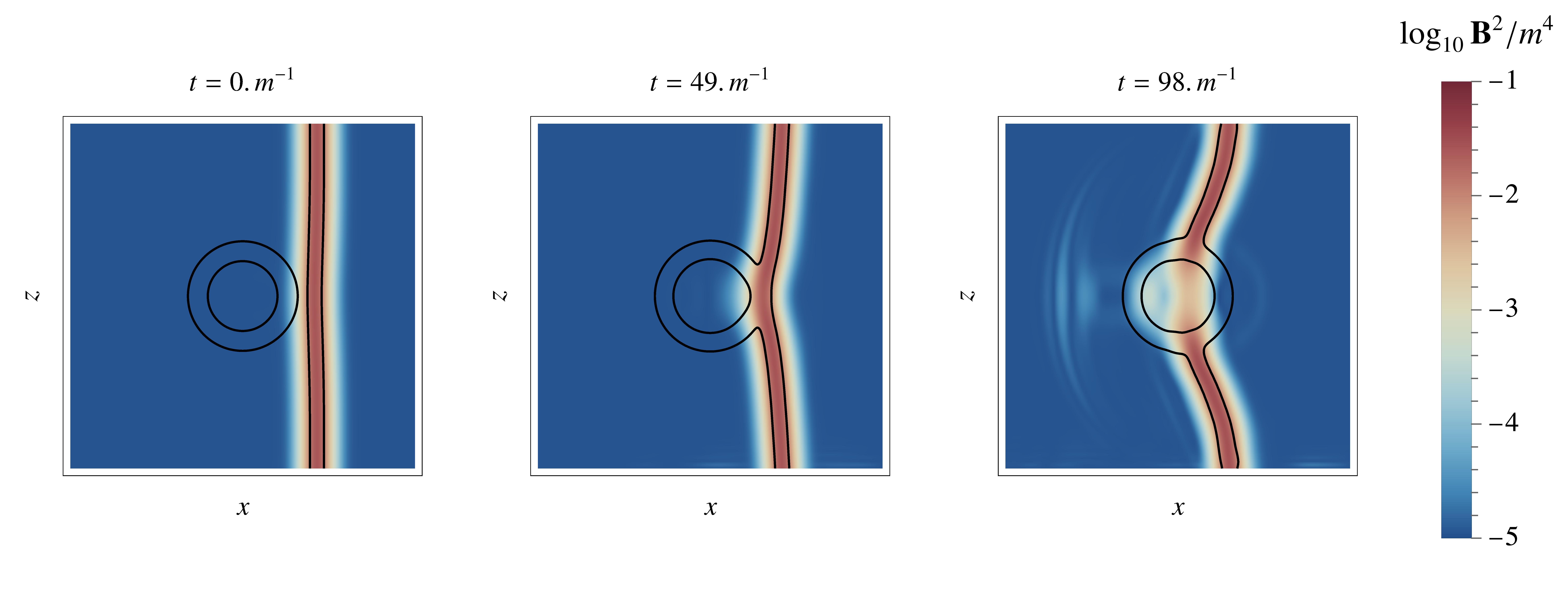}
\caption{2D snapshots of the VTM dynamics. An infinite and straight flux tube aligned with the $z$-axis (its magnetic energy density in color) is captured by a nearby $Q$-ball type bubble. Black lines are isocontours of total energy density, $|\epsilon| = 0.04\,m^{-4}$, which help to visualize the bubble.}
\label{fig:trapping_singlestring}
\end{figure*}

We first demonstrate the essence of the VTM mechanism by considering a single, infinitely long flux tube near the bubble solution. Physically, capturing a single flux tube has negligible backreaction on the bubble. However, since a numerical implementation of a large separation of scales is challenging, we neglect the vortex’s backreaction on the bubble by setting $\beta = 0$ in~(\ref{Sigma}) while evolving the adjoint field. We have confirmed that reintroducing the backreaction leaves the results qualitatively unchanged, simply adding a noise to the dynamics. Therefore, for clarity, we show the limit without backreaction.

Figure~\ref{fig:trapping_singlestring} depicts the evolution of the magnetic field energy density (in color), while the black contour marks $|\epsilon|=0.04\,m^{-4}$. As anticipated from the discussion following~(\ref{mu}), the flux tube is attracted to the bubble. Inside the bubble, the  VEV of the electromagnetic Higgs $\rho$ is lower, reducing the effective gauge boson mass and allowing for a partial spreading of magnetic field lines. This is visible in the final panel, where the trapped portion of the flux tube has lower intensity. Notice that in the above numerical simulation and in what follows we fixed couplings such that the gauge-$U(1)_{\rm EM}$ is still Higgsed in the interior of the bubble. 


\subsubsection{Saturation of the Magnetic Field: A Walk Through the Forest}

\begin{figure*}[!ht]
\centering
\includegraphics[width=1.\textwidth]{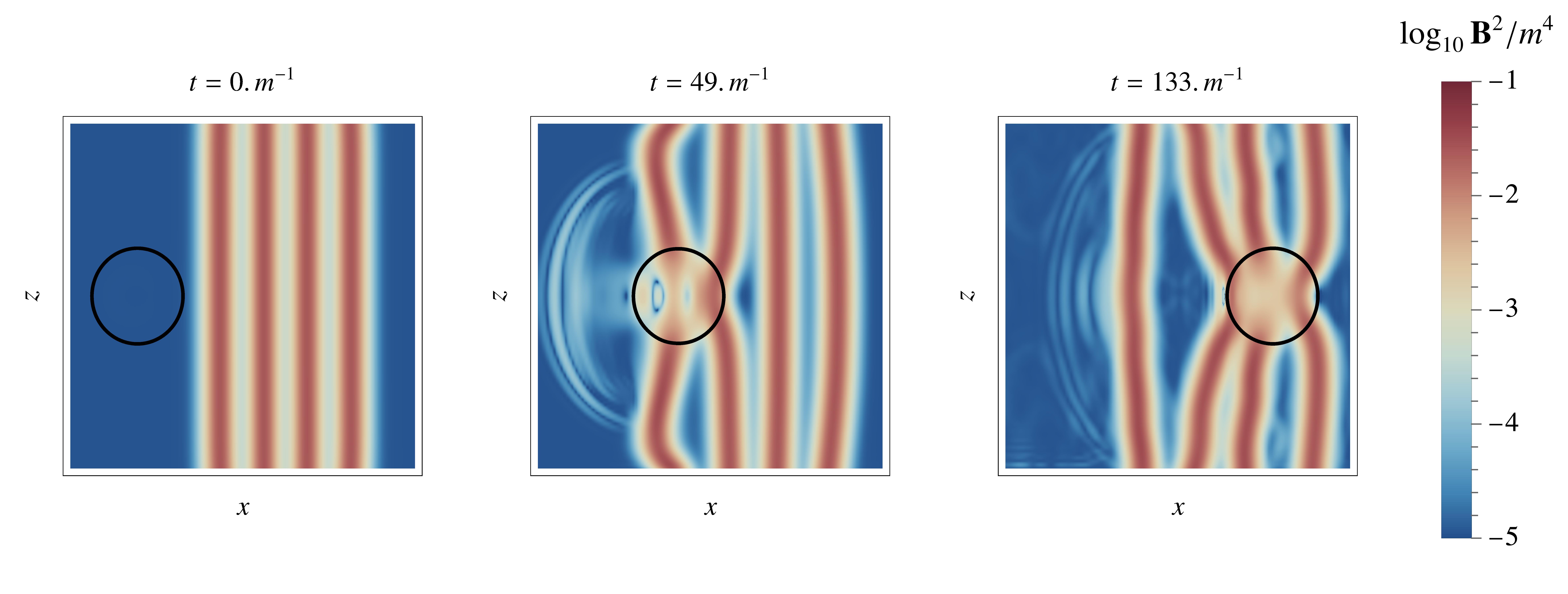}
\caption{2D snapshots of a vortex-trapping bubble passing through a “forest” of flux tubes aligned with the $x$-axis. Their magnetic energy density is shown in red-blue. The black contour denotes the Q-ball bubble, identified by the iso-contour $q = 10^{-1}\,m^{3}$. 
After some time, the bubble saturates the magnetic field it can sustain, leaving behind one of the previously trapped flux tubes.}
\label{fig:forest}
\end{figure*}

In our simulations, we set $\lambda = 1$, $v = 0.6\,m$, $\beta^2 = 0.4$, and $g=1$. While these values do not create a large hierarchy, they capture the core features of the VTM mechanism. Remarkably, for this choice, the bubble can only trap $\mathcal{O}(1)$ flux tubes.

Figure~\ref{fig:forest} shows an example where we arrange a configuration of a series of infinite straight flux tubes with total winding number 4 along the $z$-axis and boost the bubble along $x$ (first panel). The color map displays the magnetic energy density, and the black contour indicates the bubble’s location (corresponding to charge density $q=10^{-1}\,m^{3}$).

As before, the VTM mechanism traps flux tubes quite efficiently (second panel). Although a realistic black hole evolving in the galactic medium might involve more complex field configurations, our simplified setup is sufficient for illustrating the underlying physics of VTM 
phenomenon.  Once the black-hole-analog bubble approaches its maximum possible field $B_{\rm max}$, it ceases to absorb additional flux. In our example, saturation is reached after capturing roughly three flux tubes (last panel in Fig.~\ref{fig:forest}), causing the firstly-captured tube to be released. When nearing the saturation, the bubble can no longer confine more flux. Therefore,  a part of the initially trapped flux - corresponding to the first captured vortex - is ejected and left behind.


\subsubsection{Merger of Bubbles: Emission of Trapped Magnetic Field}

\begin{figure*}[!ht]
\centering
\includegraphics[width=1.\textwidth]{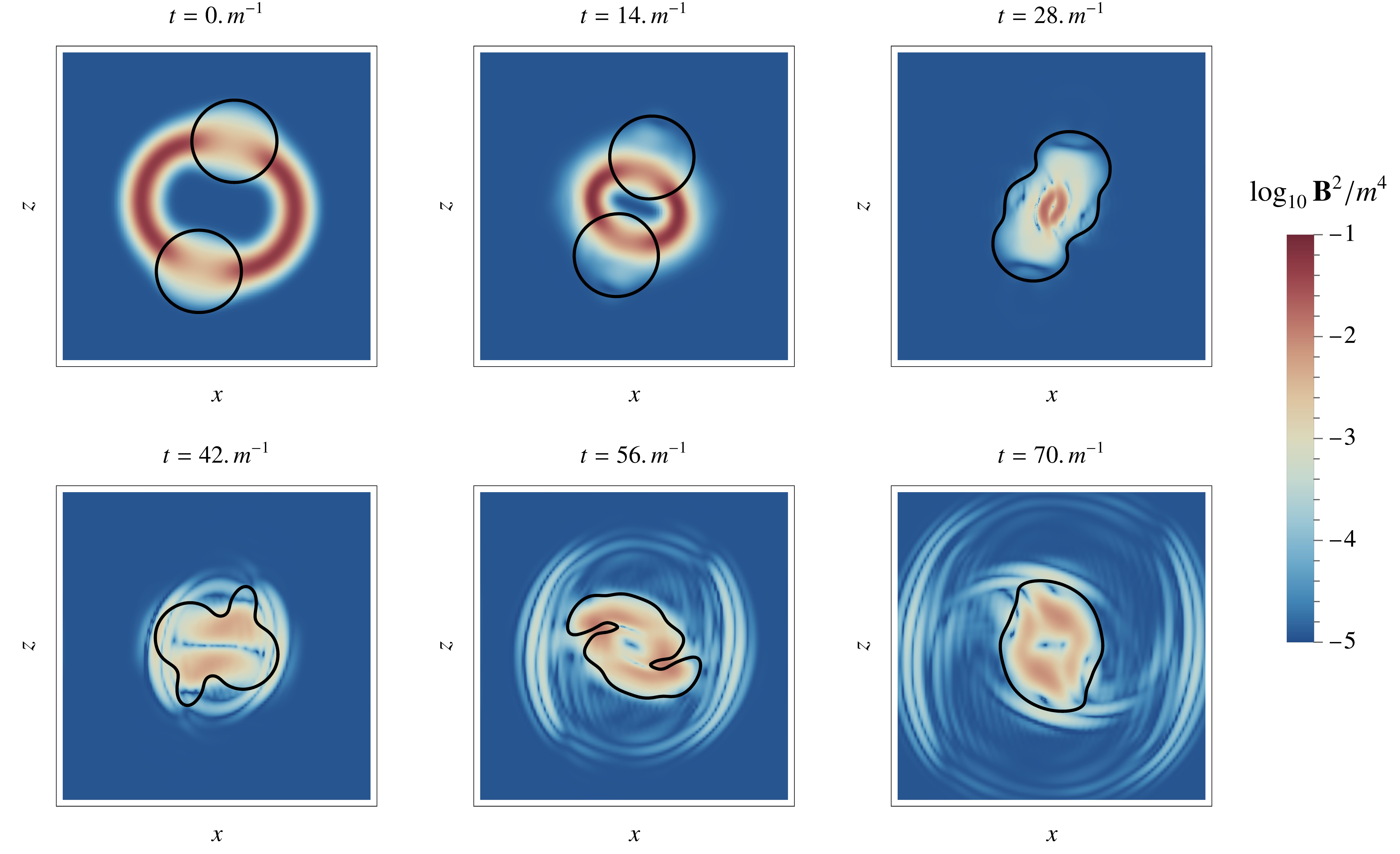}
\caption{Snapshots of two merging bubbles (black contours at $q = 10^{-1}\,m^{3}$) in the presence of a circular flux-tube loop (red). As the bubbles merge, most of the magnetic field energy is radiated away.}
\label{fig:merger}
\end{figure*}

Finally, we turn to the merger dynamics in the adjoint sector, following a setup analogous to Ref.~\cite{DvaliV2}. Two identical bubbles are placed at a separation of approximately one radius $2\,R$, with an impact parameter $R$, and boosted toward each other at $v=0.25c$. Figure~\ref{fig:merger} shows black isocontours of the bubble location $(q=10^{-1}\,m^{3}$) and the flux-tube magnetic energy density in red. The evolution of the adjoint field is analogous to the one reported in Ref.~\cite{DvaliV2} - to which we refer the interested reader for more details (see also \href{https://youtu.be/t29WUvZM-io}{URL1} and \href{https://youtu.be/zorkQSYCliU}{URL2} for visuals of the merger).

We initialize the magnetic vortex as a circular string loop piercing through both  bubbles. After numerical relaxation, it is obviously no longer perfectly circular (first panel). When the bubbles merge, the loop can annihilate, releasing most of its energy in form of electromagnetic radiation (last three panels). Only a small fraction of the original magnetic field remains inside the merged bubble.

In the gravitational analog of this scenario (i.e.\ in mergers of highly-magnetized black holes), one might encounter flux tubes of more complicated shapes and topologies, potentially leading to way more complex dynamic than the one considered here. Nevertheless, this simplified setup shows that the bulk of the magnetic energy is radiated away during the merger, consistent with the expectation. We have further verified that similar outcomes 
are provided by other geometries such as the case in which one of the bubbles is pierced by an infinitely long string perpendicular to its initial velocity. In this case, the merger dynamics leads to a highly-excited string which vibrates and emits radiation.

\section{Conclusions and outlook}\label{sec8:conclusion}

 In this paper we discussed some implications of non-zero photon mass within a consistent field-theoretic framework introduced in~\cite{ADG}. 
 The photon mass is one of the most fundamental parameters 
 of nature with an uncertain value.   
 At first glance, it may appear that the photon mass is bounded 
from above by the inverse uniformity length of a magnetic field,
$l$. According to this logic, the magnetic field of the galaxy, 
with the average strength $B_{\rm g} \sim 1\,\mu$G over the length scales 
$l_{\rm g} \gg 1\,$kpc, would provide an upper bound 
$m_{\gamma} < 3 \times 10^{-27}$ eV. One would think that 
in the opposite case the contribution from the Proca energy 
would significantly affect the galactic plasma~\cite{Galaxy1}. 

 However, as argued in~\cite{ADG}, the situation is profoundly different  since the above bounds do not take into account 
the vortex structure of the magnetic field that is intrinsic 
to the theory with non-zero photon mass.  Due to this structure, 
an uniform magnetic field of arbitrary extent, $l \gg l_{\gamma}$,
can be sustained by a large number of overlapping 
Nielsen-Olesen magnetic vortexes. 
At the same time, the Proca energy is canceled  by the winding 
of the  Higgs phase.

 This leaves us with milder bounds in the range 
 $m_{\gamma} \sim 10^{-14} - 10^{-17}$eV. 
 That is, the uniformity of the galactic magnetic field 
 over the length-scales $l_{\rm g} \gg 1\,$kpc, is fully 
 compatible with a much shorter Compton wavelength of the photon, $l_{\gamma} 
 \sim 10^{10} - 10^{13}$cm~\cite{ADG}.

     We notice that this range of the photon masses 
 matches the sizes of supermassive astrophysical black holes.
 Correspondingly, it can generate a strong magnetic field 
 via a new mechanism which we call VTM. 
  The idea is that a black hole can substantially enhance  
 its magnetic field by trapping the magnetic vortexes. 
 The physical essence of trapping is similar to
 a trapping of a cosmic string by a black hole.
  The role of the cosmic string is played by the Higgs core of 
  the magnetic vortex.   The simple estimates show that 
  the magnetic field achieved in this way can saturate the 
  upper bound (\ref{BBound}) on the magnetic capacity of a 
 black hole.  

 The actual formation mechanism of such black holes is 
 a separate question depending on the details of the cosmological scenario.  However, there exists a minimal model-independent contribution from the  exposure of a black hole to the wind of 
 the vortex network composing the magnetic field of the galaxy. 
 
 Indeed, according to the massive photon scenario of~\cite{ADG},  
 the galactic magnetic 
 field represents a densely populated forest of vortexes 
 with overlapping magnetic cores. Notice that this is not an additional assumption but a dynamically-generated structure of the galactic magnetic field imposed by a non-zero photon mass.

 These vortexes are expected to be trapped and collected by a black hole. 
 We estimate that the magnetic field
 accumulated in this way over a cosmic time is 
 (\ref{BBmax}). 
  Notice that this minimal contribution to the 
  magnetic field of a black hole is purely the effect 
  coming from the galactic magnetic field, independent 
  of the existence of any electromagnetic plasma in the Universe. 
   The presence of plasma is expected to provide further enhancement effects which require separate studies.

 The VTM effect is of direct relevance for magnetic field induced  particle acceleration scenarios such as the ones based on Fermi~\cite{fermi, bell1,bell2,catanese} or the Blandford-Znajeck~\cite{bz,bz1} mechanisms. 
 Another promising mechanism of direct relevance is provided 
by the magneto-centrifugal acceleration.  
    
The near-saturated value of the magnetic field provided by 
the VTM can have significant effects in black hole mergers.  The mergers of strongly magnetized black holes are expected to be accompanied by a highly intense 
electromagnetic radiation of wavelength $\sim R$ with the power that can potentially reach nearly-Planckian values (\ref{PowerPL}). This radiation is expected 
to carry away an order-one fraction of the magnetic energy which for the saturated value (\ref{BBound}) is comparable to the black hole mass. 

 We have provided an explicit numerical simulation of VTM effect within 
 a simple toy model (visuals can be found at the following~\href{https://youtu.be/KnxQLbEHmuU}{URL}).  In this model the role of a black hole is 
 played by a saturon bubble introduced as a 
 black hole prototype in the earlier works~\cite{Saturon2020, S4, 
DvaliV1, DvaliV2}.  While this soliton shares the relevant 
key properties with a black hole, it is easier to analyse. 
  In this model, we gave a full analytic realization of  the VTM phenomenon and also performed the numerical simulation of its real-time dynamics. 
  With the proper mapping of the parameters, the system reproduces 
  the expected features of VTM in black holes. 
   As a side remark, apart of being an useful theoretical  test prototype for a black hole,   a highly-magnetized  vacuum bubble 
would be a  rather interesting astrophysical object. 
In particular, it can mimic features of highly-magnetized  
 black holes.   
It therefore would be proper to identify such solitons in motivated extensions of the Standard Model.

The magnetodynamics  of mergers
 of VTM black holes shares some features observed in 
the simulations in analog systems of
highly-spinning saturon bubbles~\cite{SaturonV}, 
motivated by the 
alternative magnetization  scenario of~\cite{DvaliV1, DvaliV2}.

   Finally we would like to remark about some further obvious implications 
 of the considered scenario to be looked at. 
 The first is the so-called superradiance~\cite{srad1,srad2,srad3,srad4, srad5} which manifests in amplification of radiation by a rotating black hole of bosonic particles with Compton wavelengths comparable to the black hole radius.  This phenomenon has been 
 discussed for hypothetical 
 particles such as the axion-like pseudo-scalars~\cite{sradAX} or hidden sector ``dark photons"~\cite{sradDP}. 
 However, to our knowledge, this effect has never been applied to photons, since ordinarily they are considered to be massless. 
  In contrast,  the scenario considered here is ready-made for 
  this effect, since the photon mass can be in the right ballpark
  for supermassive black holes. 
  We thus expect that in resonant cases, photon superradiance can open up an additional energy loss channel for a spinning black hole. 
  This may impose restrictions on the parameters of the theory. 

 Another natural consequence of VTM would be the emission of gravitational waves from the vortexes that spin with a black hole and, in particular, from their Higgs cores. 
  Cosmic string loops are well known potential sources of gravitational waves (see,.e.g,~\cite{Vilenkin:1981bx,Vachaspati:1984gt,Blanco-Pillado:2017oxo}). The gravitational radiation from strings trapped 
   in spinning black holes has also been considered~\cite{Xing:2020ecz,Deng:2023cwh}. 
      This can have direct implications for the VTM scenario. 
  All the above effects require quantifications by precision studies. 
  
\,

\begin{acknowledgments}
\textbf{Acknowledgments} 
 We acknowledge the useful discussions and ongoing collaborations~\cite{SaturonV, DKO, MaxJuan}  with Juan Sebastian  Valbuena-Bermudez, Florian K\"uhnel and Maximilian Bachmaier. 
 We are grateful to Elisa Resconi for discussions about
 IceCube and related topics. 
 We thank Roger Blandford for illuminating communication about the magnetic fields in black holes at early stage of this work.

  Z.O. would like to thank Arnold Sommerfeld Center, Ludwig Maximilians University and Max Planck Institute for Physics for hospitality during the work on this project.

The work of G.D. was supported in part by the Humboldt Foundation under Humboldt Professorship Award, by the European Research Council Gravities Horizon Grant AO number: 850 173-6,
by the Deutsche Forschungsgemeinschaft (DFG, German Research Foundation) under Germany's Excellence Strategy - EXC-2111 - 390814868, and Germany's Excellence Strategy under Excellence Cluster Origins. The work of Z.O. was supported by Shota Rustaveli National Science Foundation of Georgia (SRNSFG). Grant: FR-23-18821 and DAAD scholarship within the
program Research Stays for University Academics and Scientists, 2024 (ID: 57693448). \\
 
 Disclaimer: Funded by the European Union. Views
and opinions expressed are however those of the authors
only and do not necessarily reflect those of the European Union or European Research Council. Neither the
European Union nor the granting authority can be held
responsible for them.
\end{acknowledgments}

\bibliography{biblio}

\end{document}